\documentclass[%
 reprint,
 amsmath,amssymb,
 aps,
]{revtex4-1}

\usepackage{graphicx}% Include figure files
\usepackage{dcolumn}% Align table columns on decimal point
\usepackage{bm}% bold math

\def\BI#1#2{\mathcal{I}_{#1}\left({#2}\right)}
\def\BK#1#2{\mathcal{K}_{#1}\left({#2}\right)}

\begin{document}

\preprint{APS/123-QED}

\title{Spontaneous motion of a camphor particle with a triangular modification from a circle}

\author{Hiroyuki Kitahata}%
 \email{kitahata@chiba-u.jp}
\affiliation{Department of Physics, Graduate School of Science, Chiba University, Chiba 263-8522, Japan
}%

\author{Yuki Koyano}
\email{koyano@cmpt.phys.tohoku.ac.jp}
\affiliation{Department of Physics, Graduate School of Science, Tohoku University, Sendai 980-8578, Japan
}%

\date{\today}% It is always \today, today,


\begin{abstract}
The spontaneous motion of a camphor particle with a slight modification from a circle is investigated. The effect of the shape on the motion is examined by the perturbation method. We introduce a slight $n$-mode modification from a circle, where the profile is described by $r = R(1 + \epsilon \cos n\theta)$ in polar coordinates. The results predict that a camphor particle with an $n=3$ mode modification from a circle, i.e., a triangular modification, moves in the direction of a corner for a small particle, while it moves in the direction of a side for a large particle. The numerical simulation results well reproduce the theoretical prediction. The present study will help understand the effect of the particle shape on spontaneous motion.
\end{abstract}

\maketitle

%\tableofcontents

\section{Introduction}

The motion of living organisms has attracted the interest of researchers of a wide variety of natural and social sciences. Physicists have been studying the mechanism of the motion of living organisms such as bacteria, bird flocks, and fish schools. From the standpoint of physics, living organisms can drive themselves by dissipating free energy under nonequilibrium conditions. In this sense, self-propelled particles are relevant examples of nonequilibrium systems. Therefore, it is also important and interesting to investigate the characteristics of self-propelled particles designated by physico-chemical systems, in which we can easily change or control the experimental parameters. There have been numerous experimental and theoretical studies on self-propelled particles and their collective behavior.\cite{Ramaswamy2010, Ohta2017, Marchetti2013, Bechinger2016,Pimienta2}

In considering the motion of such self-propelled particles, symmetric properties are essential in general. For example, the correlation between the cell shape and its migration direction was reported.\cite{Keren} Ohta and Ohkuma developed a theoretical framework by considering only the symmetric properties.\cite{Ohta2017,OhtaOhkuma} Their model has been further improved and adopted to many kinds of systems.\cite{Ebata1,Ebata2} However, the Ohta--Ohkuma model only suggests minimum criteria for the possible coupling between variables, and it is necessary to discuss actual systems to determine the coupling constants.  

A camphor--water system is a useful experimental system to investigate the coupling between the shape and the self-propulsion since we can design the shape of the particle.
This system has been intensively studied as a good example of a self-propelled particle, where the camphor particle exhibits spontaneous motion owing to the surface tension gradient generated by itself.\cite{Skey1878,Tomlinson1862,Rayleigh1889,NakataLangmuir,Hayashima,Chen,NagayamaPhysD2004,Nakata2015,BookChap2} A camphor particle floating at a water surface releases camphor molecules to the water surface, generating a surface concentration field around the particle. The surface tension gradient owing to the surface concentration field generates the force and torque acting on the camphor particle. 
The camphor particle shape is, therefore, an essential factor governing the characteristics of the spontaneous motion, because the concentration field is affected by the camphor particle shape. Moreover, the particle shape also affects the force and torque since they result from the surface tension at the particle periphery. Actually, there have been several studies focusing on the effect of the shape or deformation of the particle.\cite{NakataLangmuir,morohashi,string,rotor,rotor2,rotor3} We also investigated the motion of an elliptic camphor particle as the most fundamental modification from a circle.\cite{KitahataPRE2013,IidaPhysD2014} In our previous study, we adopted a perturbation approach and calculated the change in the bifurcation point at which the rest state is destabilized and spontaneous motion appears. The analysis predicts that an elliptic camphor particle is propelled in the direction of its minor axis, and experimental and numerical results support the theoretical prediction. The perturbation approach was applied to a camphor particle with a higher-mode modification from a circle, but we could not obtain nontrivial results.\cite{BookChap2} That is, the previous approach is only applicable to a camphor particle with an $n=2$ mode modification from a circle. We should develop analytical methods to discuss a camphor particle with a generic shape.

In the present study, we consider a camphor particle with a slight $n$-mode modification from a circle, where the shape is described by $r = R(1 + \epsilon \cos n\theta)$ in polar coordinates, and analyze the camphor particle motion in a more generic manner than in the previous work: we calculate the torque, as well as the force, when the camphor particle is moving at a constant velocity. As a result, we obtain nontrivial results for a camphor particle with an $n=3$ mode modification from a circle. 

The present results on the correlation between the motion and shape will be of value in discussing the generic correlation between the motion and dynamic deformation, for example, in a droplet system driven by a surface tension gradient.\cite{Nagai,Sumino,Annette,Pimienta,Pimienta2,Lagzi,Cejkova}
Moreover, the model has also been adopted for the motion of a chemotactic particle, which is attracted to chemicals that are released from the particle itself.\cite{Mikhailov,chemo1,chemo2} Therefore, our results can be applied not only specifically to the camphor--water system, but also to a wide variety of systems with spontaenous motion and shape deformation.

This paper is organized as follows. First, we introduce a theoretical model for the camphor particle motion in Sect.~\ref{sec2}. In Sect.~\ref{sec:analysis}, we consider the case of a circular camphor particle and then the case of a camphor particle with an $n$-mode modification from a circle. We especially focus on $n=2$ mode and $n=3$ mode modifications. Then, we check the validity of the analysis by numerical calculation in Sect.~\ref{sec:num}, and we discuss the correspondence to actual systems and the application to other systems in Sect.~\ref{sec:discussion}.

\section{Theoretical Model \label{sec2}}

The theoretical model used here is the same as the one used in the previous work.\cite{Hayashima,Chen,NagayamaPhysD2004,Nakata2015,BookChap2,KitahataPRE2013,IidaPhysD2014} The model is composed of a reaction-diffusion equation for the dynamics of the concentration of camphor molecules at a water surface, $u(\bm{r},t)$, for $\bm{r} = (x,y) \in \mathbb{R}^2$, and the Newtonian equations for the dynamics of the center position $\bm{r}_c$ and characteristic angle $\theta_c$ of the floating camphor particle. 

As the dynamics of the surface concentration of camphor molecules, $u(\bm{r},t)$, we have
\begin{align}
    \frac{\partial u}{\partial t} = D \nabla^2 u - a u + \frac{S_0}{A} \Theta(\bm{r}; \bm{r}_c, \theta_c),
\end{align}
where the first, second, and third terms in the right-hand side correspond to the diffusion at the water surface, the evaporation to the air, and the release of camphor molecules from the particle, respectively. $D$ is the diffusion coefficient and $a$ is the evaporation rate of camphor molecules from the surface. Note that the transport by the Marangoni effect due to the surface tension gradient is assumed to be included in the diffusion term,\cite{SuematsuLang2014,KitahataJCP2018,Bickel2019} where $D$ should be an ``effective'' diffusion coefficient.
Here, $S_0$ is the amount of camphor molecules released to the water surface per unit time and $A$ is the basal area of the camphor particle. The release term is explicitly described as
\begin{align}
    \Theta(\bm{r}; \bm{r}_c, \theta_c) = \left\{ \begin{array}{ll} 1, & \bm{r} \in \Omega(\bm{r}_c, \theta_c), \\ 0, & \bm{r} \notin \Omega(\bm{r}_c, \theta_c). \end{array} \right.
\end{align}
Here, $\Omega(\bm{r}_c, \theta_c)$ is the region in $\mathbb{R}^2$ that corresponds to the inside of the camphor particle, and it is described as
\begin{align}
    \Omega(\bm{r}_c, \theta_c) = \left\{ \bm{r} \middle| \mathcal{R}(-\theta_c) \left(\bm{r} - \bm{r}_c \right) \in \Omega_0 \right\},
\end{align}
where $\Omega_0$ corresponds to the region of the camphor particle when $\bm{r}_c = \textbf{0}$ and $\theta_c = 0$.
$\mathcal{R}(\theta)$ is the rotation matrix in a two-dimensional system,
\begin{align}
\mathcal{R}(\theta) = \left( \begin{array}{cc}
\cos \theta & -\sin \theta \\
\sin \theta & \cos \theta
\end{array}\right).
\end{align}
Note that the area of $\Omega_0$ is equal to $A$.

As the dynamics of the center position $\bm{r}_c$ and the characteristic angle $\theta_c$ of the camphor particle, we have
\begin{align}
    m \frac{d^2 \bm{r}_c}{dt^2} = - \eta_t \frac{d \bm{r}_c}{dt} + \bm{F},
\end{align}
\begin{align}
    I \frac{d^2 \theta_c}{dt^2} = - \eta_r \frac{d \theta_c}{dt} + N,
\end{align}
where $m$, $I$, $\eta_t$, and $\eta_r$ represent the mass, the moment of inertia, the resistance coefficient for translational motion, and that for rotational motion, respectively. The force $\bm{F}$ and torque $N$ are given as
\begin{align}
    \bm{F} =& \int_{\partial \Omega(\bm{r}_c, \theta_c)} \gamma(u(\bm{r}')) \bm{n}(\bm{r}') d\ell' \nonumber \\
    =& \iint_{\Omega(\bm{r}_c, \theta_c)} \nabla' \gamma(u(\bm{r}')) dA', \label{force_gen}
\end{align}
\begin{align}
    N =& \int_{\partial \Omega(\bm{r}_c, \theta_c)} \left( \bm{r}' - \bm{r}_c \right) \times \gamma(u(\bm{r}')) \bm{n}(\bm{r}') d\ell' \nonumber \\
    =& \iint_{\Omega(\bm{r}_c, \theta_c)} \left( \bm{r}' - \bm{r}_c \right) \times \nabla' \gamma(u(\bm{r}')) dA', \label{torque_gen}
\end{align}
where $d\ell'$ is an arc element along the periphery $\partial \Omega(\bm{r}_c,\theta_c)$ of the region $\Omega(\bm{r}_c,\theta_c)$, $\bm{n}(\bm{r})$ is a normal unit vector directing outward from the particle at the periphery $\bm{r}$, $dA'$ is an area element on $\Omega(\bm{r}_c, \theta_c)$, and $\nabla'$ is the vector differential operator with respect to $\bm{r}'$. The operator ``$\times$'' denotes $\bm{a} \times \bm{b} = a_1 b_2 - a_2 b_1$ for $\bm{a} = a_1 \bm{e}_x + a_2 \bm{e}_y$ and $\bm{b} = b_1 \bm{e}_x + b_2 \bm{e}_y$, where $\bm{e}_x$ and $\bm{e}_y$ are the unit vectors in $x$- and $y$-directions, respectively. The detailed derivations of Eqs.~\eqref{force_gen} and \eqref{torque_gen} are shown in Appendix~\ref{app}. $\gamma(u)$ is a function representing the dependence of the surface tension on the camphor molecule concentration. It is known that $\gamma(u)$ is a decreasing function\cite{camphor_surface_tension} and, for simplicity, we assume a linear relationship,
\begin{align}
    \gamma(u) = \gamma_0 - \Gamma u,
\end{align}
where $\gamma_0$ is the surface tension of pure water and $\Gamma$ is a positive constant.

Note that the camphor particle is floating on the water surface, and no external force or torque acts on it. In other words, force-free and torque-free conditions should hold. Strictly, we must solve the hydrodynamic equation with the boundary condition including the surface tension gradient.\cite{Wurger}  In our model, however, we introduce the ``effective'' force and ``effective'' torque. The effective force and effective torque are considered to correspond to the momentum and angular momentum exchanges, respectively, between the camphor particle and the water phase below the particle.

The model equations are transformed into the dimensionless form in which the length, time, concentration, and force units are $\sqrt{D/a}$, $1/a$, $S_0 / a$, and $\Gamma S_0 / \sqrt{a D}$, respectively. 
The dimensionless form used for the theoretical analysis is as follows:
\begin{align}
    \frac{\partial \tilde{u}}{\partial \tilde{t}} = \tilde{\nabla}^2 \tilde{u} - \tilde{u} + \frac{1}{\tilde{A}}\tilde{\Theta}(\tilde{\bm{r}}; \tilde{\bm{r}}_c, \theta_c), \label{concentration}
\end{align}
\begin{align}
    \tilde{\Theta}(\tilde{\bm{r}}; \tilde{\bm{r}}_c, \theta_c) = \left\{ \begin{array}{ll} 1, & \tilde{\bm{r}} \in \tilde{\Omega}(\tilde{\bm{r}}_c, \theta_c), \\ 0, & \tilde{\bm{r}} \notin \tilde{\Omega}(\tilde{\bm{r}}_c, \theta_c), \end{array} \right. \label{supply}
\end{align}
\begin{align}
    \tilde{\Omega}(\tilde{\bm{r}}_c, \theta_c) = \left\{ \tilde{\bm{r}} \middle| \mathcal{R}(-\theta_c) \left(\tilde{\bm{r}} - \tilde{\bm{r}}_c \right) \in \tilde{\Omega}_0 \right\}, \label{omega}
\end{align}
\begin{align}
    \tilde{m} \frac{d^2 \tilde{\bm{r}}_c}{d\tilde{t}^2} = - \tilde{\eta}_t \frac{d \tilde{\bm{r}}_c}{d\tilde{t}} + \tilde{\bm{F}}, \label{translation}
\end{align}
\begin{align}
    \tilde{I} \frac{d^2 \theta_c}{d\tilde{t}^2} = - \tilde{\eta}_r \frac{d \theta_c}{d\tilde{t}} + \tilde{N}, \label{rotation}
\end{align}
\begin{align}
    \tilde{\bm{F}} =& -\int_{\partial \tilde{\Omega}(\tilde{\bm{r}}_c, \theta_c)} \tilde{u}(\tilde{\bm{r}}') \bm{n}(\tilde{\bm{r}}') d \tilde{\ell}' \nonumber \\
    =& -\iint_{\tilde{\Omega}(\tilde{\bm{r}}_c, \theta_c)} \tilde{\nabla}' \tilde{u}(\tilde{\bm{r}}') d\tilde{A}', \label{force}
\end{align}
\begin{align}
    \tilde{N} =& -\int_{\partial \tilde{\Omega}(\bm{\tilde{r}}_c, \theta_c)} \left( \tilde{\bm{r}}' - \tilde{\bm{r}}_c \right) \times \tilde{u}(\tilde{\bm{r}}') \bm{n}(\tilde{\bm{r}}') d\tilde{\ell}' \nonumber \\
    =& -\iint_{\tilde{\Omega}(\tilde{\bm{r}}_c, \theta_c)} \left( \tilde{\bm{r}}' - \tilde{\bm{r}}_c \right) \times \tilde{\nabla}' \tilde{u}(\tilde{\bm{r}}') d\tilde{A}'. \label{torque}
\end{align}
Hereafter, we omit the tildes to simplify the expressions.
Here, we consider the case that the camphor particle shape for $\bm{r}_c = \bm{0}$ and $\theta_c = 0$ is expressed as
\begin{align}
    r = R \left(1 + \epsilon f(\theta)\right), \label{shape}
\end{align}
in two-dimensional polar coordinates, where $\epsilon$ is an infinitesimally small parameter. In other words, $\Omega_0$ is represented as
\begin{align}
\Omega_0 = \left\{\bm{r}(r,\theta) \mid r \leq R \left(1 + \epsilon f(\theta)\right)\right\}.
\end{align}
As the explicit form of $f(\theta)$, we consider a sinusoidal function with the wave number $n$,
\begin{align}
    f(\theta) = \cos  n \theta, \label{cosntheta}
\end{align}
where $n \in \mathbb{N}$ and $n \geq 2$. Note that we consider the modes with $n \geq 2$ because the $n=1$ mode corresponds to the translation of the center of mass, which is separately described by the equation of motion.

\section{Theoretical Analysis \label{sec:analysis}}

\subsection{Perturbation methods}

In our analysis, we adopt the perturbation method with regard to two infinitesimally small parameters: the modification amplitude from a circular shape $\epsilon$ and the translation speed $\delta$.\cite{KitahataPRE2013,IidaPhysD2014} These two infinitesimally small parameters can be introduced independently; that is, the relation between these two parameters can be chosen arbitrarily.
By considering the degree of freedom for $\theta_c$, we can set the $x$-axis to coincide with the direction of the translation velocity of the camphor particle without losing generality, and we consider the comoving frame with the camphor particle. The position of the center of mass of the camphor particle is set to the origin. In the comoving frame, whose velocity in the laboratory frame is $\delta \bm{e}_x$, Eq.~\eqref{concentration} is transformed into
\begin{align}
    \frac{\partial u}{\partial t} - \delta \bm{e}_x \cdot \nabla u  = \nabla^2 u - u + \frac{1}{A}\Theta(\bm{r}; \bm{0}, \theta_c). \label{cm_concentration}
\end{align}
Here, we calculate the steady-state concentration $u$ that satisfies
\begin{align}
    - \delta \bm{e}_x \cdot \nabla u  = \nabla^2 u - u + \frac{1}{A}\Theta(\bm{r}; \bm{0}, \theta_c). \label{cm_concentration_movingcoord}
\end{align}
The steady-state concentration field $u$ is expanded as power functions of the parameters $\delta$ and $\epsilon$ as
\begin{align}
    u = \sum_{i=0}^\infty \sum_{j=0}^\infty \delta^i \epsilon^j u_{i,j}. \label{expansion}
\end{align}
By substituting Eq.~\eqref{expansion} into Eq.~\eqref{cm_concentration}, we obtain
\begin{align}
    0 = \sum_{j = 0}^\infty \epsilon^j \left[ \nabla^2 u_{0,j} - u_{0,j}\right] + \frac{1}{A}\Theta(\bm{r};\bm{0},\theta_c) \label{st_cm_concentration}
\end{align}
for the order of $\delta^0$ and 
\begin{align}
    -\sum_{j = 0}^\infty \epsilon^j \bm{e}_x \cdot \nabla u_{i-1,j} = \sum_{j = 0}^\infty \epsilon^j \left[ \nabla^2 u_{i,j} - u_{i,j}\right] \label{st_cm_concentration2}
\end{align}
for the order of $\delta^i$ ($i \geq 1)$.

The boundary of the region of the camphor particle $\partial \Omega(\bm{0}, \theta_c)$ is explicitly represented in polar coordinates as
\begin{align}
    r =& R\left[1 + \epsilon F(\theta)\right] \nonumber \\
    =& R\left[1 + \epsilon f(\theta - \theta_c)\right] \nonumber \\
    =& R\left[1 + \epsilon \cos n(\theta- \theta_c) \right]. \label{eq_shape}
\end{align}
Note that $\theta_c$ has $(2\pi/n)$-periodicity.
Considering that $\Theta(\bm{r}; \bm{0}, \theta_c)$ has a discontinuity at the boundary, the steady-state concentration fields $u$ are obtained separately in the regions inside and outside of $\Omega(\bm{0}, \theta_c)$, which are denoted as $u^{(i)}$ and $u^{(o)}$, respectively.
Continuity conditions between $u^{(i)}$ and $u^{(o)}$ are required; $u$ should be $C^1$-class because the discontinuity of $\Theta(\bm{r}; \bm{0}, \theta_c)$ in Eq.~\eqref{st_cm_concentration} should be compensated by $\nabla^2 u$, and therefore, the second derivative of $u$ should have discontinuity. 
The continuity conditions are explicitly expressed as
\begin{align}
    u^{(i)}(R(1+\epsilon F(\theta)),\theta) = u^{(o)}(R(1+\epsilon F(\theta)),\theta)
\end{align}
and
\begin{align}
    \left.\frac{\partial u^{(i)}}{\partial r}\right|_{r =R(1+\epsilon F(\theta))} = 
    \left.\frac{\partial u^{(o)}}{\partial r}\right|_{r =R(1+\epsilon F(\theta))}.
\end{align}
By describing $u^{(i)}$ and $u^{(o)}$ by the expansions
\begin{align}
u^{(i)} = \sum_{i=0}^\infty \sum_{j=0}^\infty \delta^i \epsilon^j u^{(i)}_{i,j}, \label{ui}
\end{align}
\begin{align}
u^{(o)} = \sum_{i=0}^\infty \sum_{j=0}^\infty \delta^i \epsilon^j u^{(o)}_{i,j}, \label{uo}
\end{align}
these conditions are expressed as
\begin{align}
    u^{(i)}_{i,0}(R, \theta) =  u^{(o)}_{i,0}(R ,\theta), \label{cont_cond0a}
\end{align}
\begin{equation}
\left. \frac{\partial u^{(i)}_{i,0}}{\partial r} \right|_{r = R} = \left. \frac{\partial u^{(o)}_{i,0}}{\partial r} \right|_{r = R} \label{cont_cond0b}
\end{equation}
for the order of $\epsilon^0 \delta^i$ ($i \geq 0$) and
\begin{align}
    &\sum_{k=0}^\nu \frac{1}{k!} \left[R F(\theta)\right]^{k} \left.\frac{\partial^k u_{i,\nu-k}^{(i)}}{\partial r^k} \right|_{r=R} \nonumber \\ &= \sum_{k=0}^\nu \frac{1}{k!} \left[R F(\theta)\right]^k \left.\frac{\partial^k u_{i,\nu-k}^{(o)}}{\partial r^k} \right|_{r=R},
\end{align}
\begin{align}
    &\sum_{k=0}^\nu \frac{1}{k!} \left[R F(\theta)\right]^{k} \left.\frac{\partial^{k+1} u_{i,\nu-k}^{(i)}}{\partial r^{k+1}} \right|_{r=R} \nonumber \\ &= \sum_{k=0}^\nu \frac{1}{k!} \left[R F(\theta)\right]^{k} \left.\frac{\partial^{k+1} u_{i,\nu-k}^{(o)}}{\partial r^{k+1}} \right|_{r=R}
\end{align}
for the order of $\epsilon^\nu \delta^i$ ($\nu \geq 1$, $i \geq 0$). In the present study, we consider the terms up to $\delta^3$ and $\epsilon^1$.

Note that the area $A$ of $\Omega_0$ is obtained as
\begin{align}
    A = \pi R^2 \left( 1 + \frac{1}{2}\epsilon^2 \right)
\end{align}
for the shape in Eq.~\eqref{eq_shape} ($n \geq 2$). Here, the term proportional to $\epsilon^2$ can be neglected since we only consider the concentration up to the order of $\epsilon$.

\subsection{Circular camphor particle}

First, we consider the case with no modification from a circle. This corresponds to the concentration field $u_{i,0}$.
$u_{0,0}$ is obtained in polar coordinates\cite{KitahataPRE2013,IidaPhysD2014} as
\begin{align}
    u_{0,0}^{(i)} = \frac{1}{\pi R^2}\left[ 1 - R \BK1{R} \BI0{r} \right], 
\end{align}
\begin{align}
    u_{0,0}^{(o)} = \frac{1}{\pi R^2} \left[ R \BI1{R} \BK0{r} \right],
\end{align}
where $\BI{n}{\cdot}$ and $\BK{n}{\cdot}$ are the modified Bessel functions of first and second kinds with the order of $n$, respectively.
The steady-state concentration field is obtained by solving Eq.~\eqref{st_cm_concentration2} successively. The explicit forms for $u_{i,0}^{(i)}$ and $u_{i,0}^{(o)}$ are shown in Appendix~\ref{appB}.
Using the descriptions, the force and torque in Eqs.~\eqref{force} and \eqref{torque} are calculated as
\begin{align}
    \bm{F}^{(0)} = \left[ f_{1,0} \delta - f_{3,0} \delta^3 \right] \bm{e}_x,
\end{align}
\begin{align}
    N^{(0)} = 0,
\end{align}
where
\begin{align}
    f_{1,0} = \frac{R^2}{4}\left[ \BI0{R}\BK0{R} - \BI2{R} \BK2{R} \right], \label{f10}
\end{align}
\begin{align}
    f_{3,0} =& \frac{R^4}{32} \left[ - \BI0{R}\BK0{R} + \frac{2}{R^2} \BI1{R}\BK1{R}  + \BI2{R}\BK2{R} \right]. \label{f30}
\end{align}
Here, $f_{1,0} > 0$ and $f_{3,0} > 0$ for all $R > 0$.

From these results, the dynamical systems for a circular camphor particle motion can be written as
\begin{align}
m \frac{d\bm{v}_c}{dt} =& -\eta_t \bm{v}_c + \bm{F} \nonumber \\
=& -\eta \bm{v}_ c + f_{1,0} \bm{v}_c - f_{3,0} \left| \bm{v}_c \right|^2 \bm{v}_c
\end{align}
for the camphor particle velocity $\bm{v}_c$ with a small absolute value.
This suggests that a circular camphor particle moves if $\eta_t < f_{1,0}$ and that it stops if $\eta_t > f_{1,0}$. Considering $f_{3,0} > 0$, this can be considered as a supercritical pitchfork bifurcation or drift bifurcation in dynamical systems. In other words, the bifurcation point for the translational motion is given by the condition $\eta_t - \eta_t^{(bp)} = 0$, where
\begin{align}
\eta_t^{(bp)} = f_{1,0}. \label{etatbp}
\end{align}
These results are consistent with the ones shown in the previous work.\cite{KitahataPRE2013,IidaPhysD2014,BookChap2}

\subsection{Camphor particle with $n$-mode shape}

In the case of a camphor particle with a slight $n$-mode modification from a circle as shown in Eq.~\eqref{eq_shape}, the expanded concentrations can be explicitly given as shown in Appendix~\ref{appB}.
Using the explicit descriptions, the force and torque in Eqs.~\eqref{force} and \eqref{torque} are calculated.

In the case of $n=2$, we obtain the force $\bm{F}^{(2)}$ and torque $N^{(2)}$ from Eqs.~\eqref{force}, \eqref{torque}, and \eqref{ui} with the explicit forms in Appendix~\ref{appB}:
\begin{align}
    \bm{F}^{(2)} =& \left[ (f_{1,0} - f^{(2)}_{1,1} \epsilon \cos 2\theta_c) \delta  - (f_{3,0} - 2f^{(2)}_{3,1} \epsilon \cos 2\theta_c )\delta^3 \right] \bm{e}_x \nonumber \\
    &- \left[ (f^{(2)}_{1,1} \epsilon \sin2\theta_c) \delta - (f^{(2)}_{3,1} \epsilon \sin2\theta_c) \delta^3 \right] \bm{e}_y,
\end{align}
\begin{align}
    N^{(2)} =  g^{(2)}_{2,1} \epsilon \delta^2 \sin 2\theta_c,
\end{align}
where
\begin{align}
    f^{(2)}_{1,1} = \frac{R^2}{2}\left[ \BI1{R}\BK1{R} - \BI2{R} \BK2{R} \right] ,
\end{align}
\begin{align}
    f^{(2)}_{3,1} =& \frac{R^4}{96} \left[ 3\BI0{R}\BK0{R} - 4 \BI1{R}\BK1{R} + \BI2{R}\BK2{R} \right],
\end{align}
\begin{align}
    g^{(2)}_{2,1} =& \frac{R^4}{32} \left[ 2\BI0{R}\BK0{R} - \BI1{R}\BK1{R} \right. \nonumber \\
    & \left. \quad  - 2\BI2{R}\BK2{R} + \BI3{R}\BK3{R} \right].
\end{align}
Here, $f^{(2)}_{1,1} > 0$, $f^{(2)}_{3,1} > 0$, and $g^{(2)}_{2,1} > 0$ for all $R > 0$. Note that $f_{1,0}$ and $f_{3,0}$ are respectively defined in Eqs.~\eqref{f10} and \eqref{f30}.

These results suggest that the force is directed along the $x$-axis direction only for $\theta_c = 0$ or $\pi/2$ considering that $\theta_c$ has $\pi$-periodicity. In such cases, the bifurcation point for the translational motion is given by the condition $\eta_t - \eta_t^{(bp,2)} = 0$, where
\begin{align}
\eta_t^{(bp,2)} = f_{1,0} - \epsilon f^{(2)}_{1,1} \cos 2\theta_c.
\end{align}
The maximum value for $\eta_t^{(bp,2)}$ is achieved when $\cos 2\theta_c = -1$, that is, $\theta_c = \pi/2$. Therefore, the bifurcation point for the translational motion is the greatest in the minor-axis direction. This means that the translational motion in the minor-axis direction is realized for a larger resistance coefficient, that is, an elliptic camphor particle tends to move in its minor-axis direction, as in the discussion in our previous papers.\cite{KitahataPRE2013,IidaPhysD2014,BookChap2}

From the present results, we can also discuss the torque acting on a camphor particle moving at a constant velocity. The torque depends on the characteristic angle $\theta_c$, namely, the torque is proportional to $\sin 2\theta_c$. Considering that the torque is acting on an elliptic camphor particle moving in the $x$-axis direction if and only if $\theta_c \neq 0, \pi/2$, the characteristic angle $\theta_c$ should be $\theta_c = 0$ or $\pi/2$ for the solution with constant $\theta_c$. By considering the sign of the torque, it is concluded that the solutions with $\theta_c = 0$ and $\pi/2$ are unstable and stable, respectively.
In other words, the direction of the elliptic camphor particle motion should converge to its minor-axis direction.

\begin{figure}
    \centering
    \includegraphics{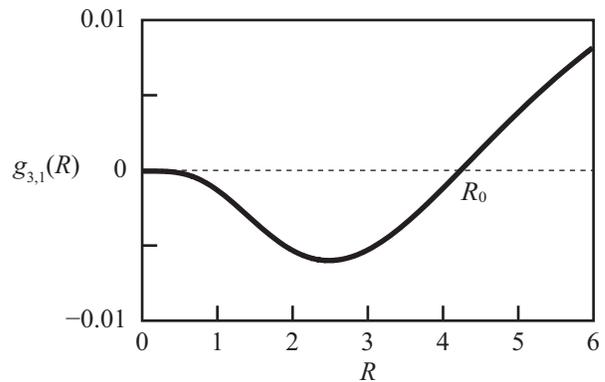}
    \caption{Plot of the coefficient $g_{3,1}^{(3)}$ against the particle radius $R$. $g_{3,1}^{(3)}$ is positive and negative for $R > R_0$ and $R < R_0$, respectively, where $R_0 \simeq 4.23024$.}
    \label{fig1}
\end{figure}

Next, we consider the case of $n=3$. We obtain the force $\bm{F}^{(3)}$ and torque $N^{(3)}$ from Eqs.~\eqref{force}, \eqref{torque}, and \eqref{ui} with the explicit forms in Appendix~\ref{appB}:
\begin{align}
    \bm{F}^{(3)} =& \left[ f_{1,0} \delta - f_{3,0} \delta^3 \right] \bm{e}_x ,
\end{align}
\begin{align}
    N^{(3)} =  g^{(3)}_{3,1} \epsilon \delta^3 \sin 3\theta_c, \label{torque3}
\end{align}
where
\begin{align}
    g^{(3)}_{3,1} =& \frac{R^5}{384} \left[ -3\BI0{R}\BK0{R} + 3\BI1{R}\BK1{R} \right. \nonumber \\
    & \left. \quad  + 2\BI2{R}\BK2{R} - 3\BI3{R}\BK3{R}  + \BI4{R}\BK4{R} \right].
\end{align}
Note that $f_{1,0}$ and $f_{3,0}$ are respectively defined in Eqs.~\eqref{f10} and \eqref{f30}.
These results suggest that the bifurcation point for the translational motion does not depend on the direction of the characteristic angle $\theta_c$ of the camphor particle. However, once the particle begins to move, the torque acts on the particle. The direction of the torque depends on the sign of $g^{(3)}_{3,1}$. Interestingly, the sign of $g^{(3)}_{3,1}$ depends on the particle radius $R$. In Fig.~\ref{fig1}, the plot of $g^{(3)}_{3,1}$ against $R$ is shown, in which $g^{(3)}_{3,1}$ is negative and positive for $R < R_0$ and $R > R_0$, respectively. $R_0$ is numerically obtained as
\begin{align}
    R_0 \simeq 4.23024.
\end{align}
This result suggests that a smaller particle moves in the direction of a corner, a more convex part, while a larger particle moves in the direction of a side, a flatter part.  

\section{Numerical Calculation \label{sec:num}}

A numerical calculation was performed to confirm the theoretical results. Equations~\eqref{concentration} to \eqref{cosntheta} were used for the numerical calculation, though Eqs.~\eqref{supply}, \eqref{force}, and \eqref{torque} were slightly changed as follows for stable numerical calculation:
\begin{align}
    \Theta(\bm{r}; \bm{r}_c, \theta_c) = \frac{1}{2} \left[1 + \tanh \frac{\left| \bm{r} - \bm{r}_c \right| - f(\phi - \theta_c)}{\Delta w} \right],
\end{align}
\begin{align}
    \bm{F} =& -\int_{\mathbb{R}^2} \left(\nabla' u(\bm{r}')\right) \Theta(\bm{r'}; \bm{r}_c, \theta_c) dA', 
\end{align}
\begin{align}
    N =& -\int_{\mathbb{R}^2} \left( \bm{r}' - \bm{r}_c \right) \times \left(\nabla' u(\bm{r}')\right) \Theta(\bm{r'}; \bm{r}_c, \theta_c) dA',
\end{align}
where $\Delta w$ is a smoothing parameter and $\phi$ is the angle of $\bm{r} - \bm{r}_c$ from the $x$-axis. We changed $R$ as a parameter. 
The periodic boundary condition was adopted for the concentration field $u$, which was defined in a square with a side of $20R$.
We used the Euler method for the time development and the explicit method for the spatial derivative. The space and time steps were set to $\Delta x = R / 40$ and $\Delta t = 10^{-4} \times R^2$, respectively. The parameters $m$ and $I$ were set as
\begin{align}
m = \rho A = \rho \pi R^2 \left(1 + \frac{\epsilon^2}{2} \right),
\end{align}
\begin{align}
I = \frac{\rho \pi R^4}{4} \left( 1 + 3 \epsilon^2 + \frac{3}{8} \epsilon^4 \right).
\end{align}
Here, we set $\rho = 10^{-3}$.
The friction coefficient for the translational motion was determined from the bifurcation point for a circular particle,
\begin{align}
    \eta_t = 0.9 \eta_t^{(bp)},
\end{align}
where $\eta_t^{(bp)}$ is defined in Eq.~\eqref{etatbp}.
The friction coefficient for the rotational motion was set so that spontaneous rotation would not occur. For the rotational motion without translational motion, the bifurcation point is given as $\eta_r = \eta_r^{(bp,n)}$ for a camphor particle with an $n$-mode modification from a circle,\cite{BookChap2} where
\begin{align}
      \eta_r^{(bp,n)} = \frac{\epsilon^2 n R^4}{4} \left[ \mathcal{I}_{n-1}(R) \mathcal{K}_{n-1}(R) - \mathcal{I}_{n+1}(R) \mathcal{K}_{n+1}(R) \right], 
\end{align}
and we set
\begin{align}
    \eta_r = 1.1 \eta_r^{(bp,n)}.
\end{align}
The parameter representing the modification amplitude was set as
$\epsilon = 0.1$.

\begin{figure}
    \centering
    \includegraphics{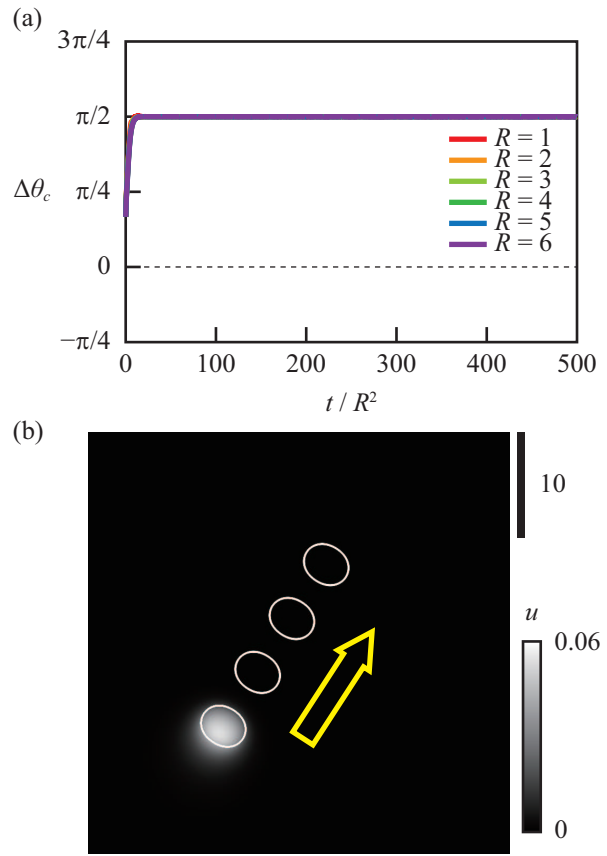}
    \caption{Numerical results in the case of an $n=2$ mode modification. (a) Time series of $\Delta \theta_c$ for $R = 1$ (red), $2$ (orange), $3$ (yellow green), $4$ (green), $5$ (blue), and $6$ (purple). $\Delta \theta_c$ is plotted against $t/R^2$. For all $R$, $\Delta \theta_c$ converges to $\pi/2$. (b) Concentration field $u$ for $R=2$ after a sufficiently long time from the start of simulation. The positions and directions of the camphor particle with a time interval of 20 are shown with the white curves. The direction of motion is indicated with a yellow arrow. A scale bar is shown on the right.}
    \label{fig:sim2}
\end{figure}

The results of the numerical calculation in the case of a two-mode modification ($n=2$) are shown in Fig.~\ref{fig:sim2}.
In order to clearly show the relationship between the direction of motion and the shape, the angle difference between the direction of velocity $\phi_c$ and the characteristic angle $\theta_c$ is defined as
\begin{align}
    \Delta \theta_c = \theta_c - \phi_c.
\end{align}
Here, $\phi_c$ is defined as
\begin{align}
    \bm{v}_c = \left| \bm{v}_c \right| \left( \begin{array}{c} \cos\phi_c \\ \sin \phi_c \end{array}\right).
\end{align}
Note that $\Delta \theta_c$ is set in the range of
\begin{align}
    - \frac{\pi}{2n} < \Delta \theta_c \leq \frac{3\pi}{2n},
\end{align}
reflecting the $(2 \pi / n)$-periodicity of $\theta_c$.
The time evolution of $\Delta \theta_c$ for $n=2$ is shown in Fig.~\ref{fig:sim2}(a), which illustrates that $\Delta \theta_c$ eventually converges to $\pi/2$. This means that the particle moves along its minor-axis direction.
We also show a snapshot of the camphor concentration field $u$ together with the particle shape and the motion direction after a sufficiently long time in Fig.~\ref{fig:sim2}(b). The camphor particle moves along its minor-axis direction.

\begin{figure}
    \centering
    \includegraphics{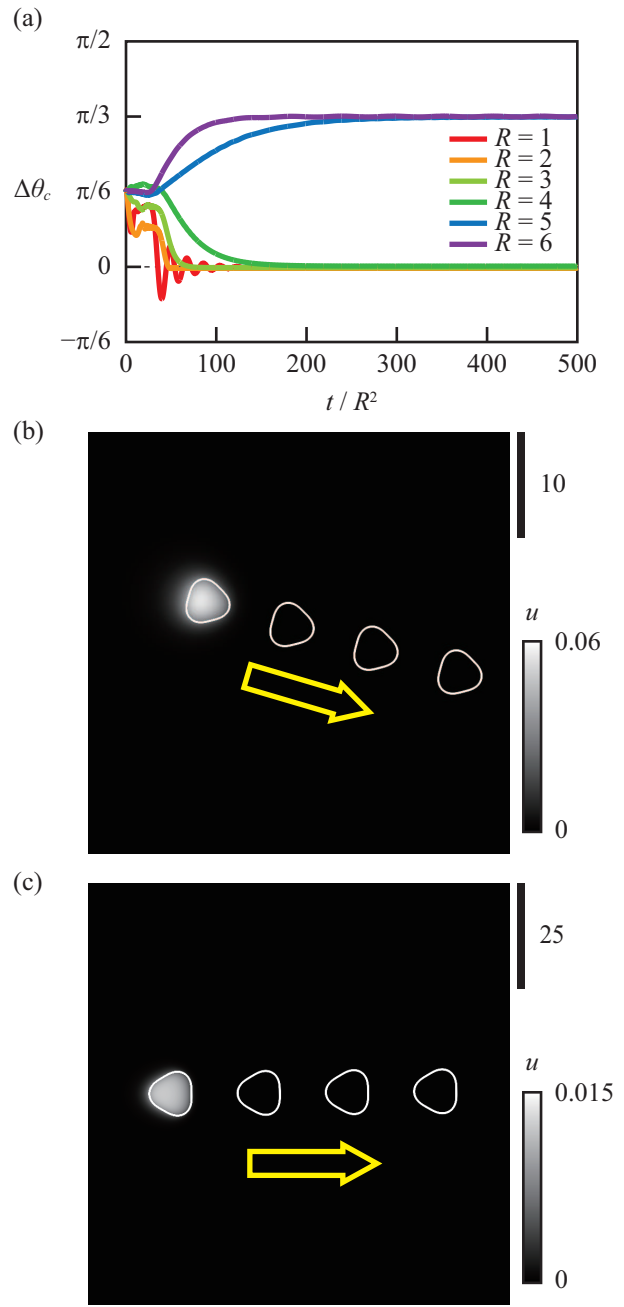}
    \caption{Numerical results in the case of an $n=3$ mode modification. (a) Time series of $\Delta \theta_c$ for $R = 1$ (red), $2$ (orange), $3$ (yellow green), $4$ (green), $5$ (blue), and $6$ (purple). $\Delta \theta_c$ is plotted against $t/R^2$. $\Delta \theta_c$ converges to $0$ for $R=1$, $2$, $3$, and $4$, while it converges to $\pi/3$ for $R=5$ and $6$. (b, c) Concentration field $u$ for (b) $R=2$ and (c) $R=5$ after a sufficiently long time from the start of simulation. The positions and directions of the camphor particle with a time interval of (b) 40 and (c) 250 are shown with the white curves. The directions of motion are indicated with yellow arrows. Respective scale bars are shown on the right.}
    \label{fig:sim3}
\end{figure}

Figure~\ref{fig:sim3} shows the numerical results in the case of three-mode modification ($n=3$). In Fig.~\ref{fig:sim3}(a), the time evolution of $\Delta \theta_c$ for $n=3$ is plotted for each $R$. $\Delta \theta_c$ converges to $0$ for $R = 1$, $2$, $3$, and $4$, while it converges to $\pi/3$ for $R = 5$ and $6$. In other words, the triangular camphor particle moves in the direction of a corner for $R = 1$, $2$, $3$, and $4$, while it moved in the direction of a side for $R = 5$ and $6$. These results well correspond to the theoretical prediction that a triangular particle moves in the direction of a corner and a side for $R < R_0$ and $R > R_0$, respectively, where $R_0 \simeq 4.23024$. Compared with the case of $n=2$, the change in $\Delta \theta_c$ is slower since the torque is proportional to $\delta^3$ for $n=3$, while it is proportional to $\delta^2$ for $n=2$. Figures~\ref{fig:sim3}(b) and \ref{fig:sim3}(c) show snapshots of the camphor concentration field $u$ together with the particle shape and the motion direction for $R = 2$ [Fig.~\ref{fig:sim3}(b)] and $R=5$ [Fig.~\ref{fig:sim3}(c)] after a sufficiently long time.

\section{Discussion \label{sec:discussion}}

In the present study, we performed theoretical analyses for the camphor particle motion with an infinitesimally small $n$-mode modification from a circle. We calculated the concentration field by the perturbation method up to the third order of velocity $\delta$ and the first order of the modification amplitude $\epsilon$. Using the calculated concentration field, we obtained the force and torque that act on the particle. In our previous study,\cite{KitahataPRE2013,IidaPhysD2014,BookChap2} we could calculate the force in the case that the velocity is in the direction of either the major or minor axis for a camphor particle with an $n=2$ mode modification. However, we could not determine the preferred moving direction for a camphor particle with a three- or higher-mode modification from a circle using the same framework since we did not calculate the torque for a camphor particle exhibiting translational motion. In the present study, we discussed the time evolution of the characteristic angle $\theta_c$ by calculating the torque for the general configuration, and thus could derive the preferred moving direction for a camphor particle with an $n=3$ mode modification. 

The coefficient $g^{(3)}_{3,1}$ used to determine the direction of the torque in Eq.~\eqref{torque3} originates from the term with $u_{3,0}$, which corresponds to the camphor concentration without shape modification. The torque is given by the coupling between $u_{3,0}$ and the modified shape of the camphor particle with the order of $\epsilon$. The explicit forms for $u_{3,0}^{(i)}$ and $u_{3,0}^{(o)}$ are respectively given in Eqs.~\eqref{u30i} and \eqref{u30o}, which are composed of a term proportional to $\cos \theta$ and one proportional to $\cos 3\theta$. The latter term is essential to calculate $g^{(3)}_{3,1}$. This means that a camphor concentration field affected by translational motion can induce a torque acting on the particle, and the torque changes the characteristic angle of the camphor particle. So far, a more detailed physical description for the preferred moving direction has not yet been found, and it remains as future work. 

For a camphor particle with a modification of $n \ge 4$ from a circle, we cannot determine the preferred moving direction by calculating only the concentration field with respect to the third order of the velocity $\delta$. It may be possible to calculate it if the concentration field is calculated up to the $n$th order of $\delta$, and we expect that the torque is on the order of $\epsilon \delta^n$. Considering that the present theory is applicable only in the regime close to the bifurcation point, the term of $\delta^n$ with larger $n$ becomes so small that the torque can practically rotate the particle very slowly. In this sense, the discussion for the preferred moving direction for larger $n$ might become meaningless.

Here, we consider the correspondence between the theoretical and experimental results. The effective diffusion coefficient is estimated as $D \simeq 4 \times 10^{-3} \, \textrm{m}^2 \, \textrm{s}^{-1}$ and the evaporation rate is estimated as $a = 2 \times 10^{-2} \, \textrm{s}^{-1}$.\cite{SuematsuLang2014} Therefore, the diffusion length $\ell_d$ is calculated as
\begin{align}
\ell_d = \sqrt{D/a} \simeq 0.4 \, \textrm{m},
\end{align}
which is the length unit in the dimensionless form. The size of a camphor particle is typically around $R_\textrm{exp} \sim 10 \, \textrm{mm} = 0.01 \, \textrm{m}$ and thus the camphor particle radius $R = R_\textrm{exp}/\ell_d$ is smaller than 1. This suggests that a camphor particle with an $n=3$ mode modification from a circle moves in the direction of a corner. The preliminary experiments using filter paper with camphor show motion in the direction of a corner (data not shown), which agrees with the theoretical prediction. To realize a camphor particle with $R > R_0$, we must prepare a large camphor particle or reduce the diffusion length by using a viscous fluid or by increasing the evaporation rate. The confirmation of the theoretical prediction by experiments remains as future work.

The motion of a pointlike camphor particle confined in a circular region was recently analyzed.\cite{Koyano2019} It would be interesting to discuss the coupling between the particle shape and boundary shape with respect to the symmetric properties. The interaction among particles has also been studied.\cite{Soh,Soh2,SuematsuJPSJ,Nishimori2017,SuematsuJPSJ2019} We discussed the interaction between the two elliptic camphor particles whose center positions are fixed.\cite{EiPhysD2018} As a natural extension of these previous studies, the dynamics of mobile particles that interact with each other depending on their characteristic direction, such as a nematic interaction, is an interesting problem. Using a multiple-camphor-particle system with a modification from a circle, we believe we can realize such a system.

It is also known that a pentanol droplet on an almost saturated pentanol aqueous solution spontaneously moves with a banana-shaped deformation.\cite{Nagai} The mechanism of the motion is almost the same as that for a camphor particle at the water surface. Other systems of moving droplets with deformation, which are driven by the surface tension gradient, have also been reported.\cite{Pimienta,Pimienta2,Lagzi,Cejkova} In the present study, we establish how to calculate the effect of the shape on the motion. For the droplet system, the motion can affect the deformation, and thus the motion and deformation are coupled with each other. The present work can be further extended to clarify the mechanism of the coupling between motion and deformation.

\section{Summary \label{sec:summary}}

In the present study, we performed theoretical analysis of the dynamics of a camphor particle with a slight $n$-mode modification from a circle by the perturbation method. For the $n=2$ mode modification, the theoretical discussion reveals that the particle moves in its minor-axis direction. For the $n=3$ mode modification, the theoretical discussion indicates that a smaller camphor particle moves in the direction of a corner, while a larger camphor particle moves in the direction of a side. Our numerical calculation corresponds to the theoretical results. The present study will help understand the effect of the particle shape on spontaneous motion.

\section*{Acknowledgments}
The authors acknowledge Masaharu Nagayama and Yutaka Sumino for their helpful discussion. This work was supported by JSPS KAKENHI Grant Nos. JP16H03949, JP19K03765, JP19J00365, and JP20K14370, and the Cooperative Research Program of ``Network Joint Research Center for Materials and Devices'' (Nos.~20194006, 20191030, 20204004, and 20201023).
This work was also supported by JSPS and PAN under the Japan--Poland Research Cooperative Program ``Spatio-temporal patterns of elements driven by self-generated, geometrically constrained flows'' and "Complex spatio-temporal structures emerging from interacting self-propelled particles" (No.~JPJSBP120204602).

\appendix

\section{Derivations of Eqs.~\eqref{force_gen} and \eqref{torque_gen} \label{app}}

Here, we derive the last line in Eqs.~\eqref{force_gen} and \eqref{torque_gen}. We set the polar coordinates $(r', \theta')$, so that the arbitrary position $\bm{r}'$ is described as
\begin{align}
    \bm{r}' = \bm{q}(\theta') = R\left[1 + F(\theta') \right] \bm{e}(\theta'),
\end{align}
where $\bm{e}(\theta')$ is a unit vector in the direction of $\theta'$.
For the force in Eq.~\eqref{force_gen}, we obtain
\begin{align}
    F_i =& \int_{\partial \Omega(\bm{r}_c, \theta_c)} \gamma(u(\bm{r}')) n_{i}(\bm{r}') d \ell' \nonumber \\
=& \int_0^{2\pi} \gamma(u(\bm{q}(\theta'))) \varepsilon_{ij} \frac{d q_j}{d\theta'} d\theta' \nonumber \\
=& \left[ \gamma(u(\bm{q}(\theta))) \varepsilon_{ij} q_j(\theta) \right]_0^{2\pi} \nonumber \\
&- \int_0^{2\pi} \left(\frac{d}{d\theta'} \gamma(u(\bm{q}(\theta')))\right)  \varepsilon_{ij} q_j(\theta') d\theta' \nonumber \\
=& -  \int_0^{2\pi}  \frac{\partial \gamma}{\partial q_k} \frac{d q_k}{d\theta'}  \varepsilon_{ij} q_j(\theta') d\theta' \nonumber \\
=& - \int_{\partial \Omega(\bm{r}_c, \theta_c)} \frac{\partial \gamma}{\partial q_k} \varepsilon_{ij} q_j d q_k \nonumber \\ 
=& - \iint_{\Omega(\bm{r}_c, \theta_c)} \varepsilon_{mk} \frac{\partial}{\partial r'_m}\left( \frac{\partial \gamma}{\partial r'_k}  \varepsilon_{ij} r'_j \right) d A' \nonumber \\
=& -  \iint_{\Omega(\bm{r}_c, \theta_c)} \varepsilon_{ij} \varepsilon_{jk}  \frac{\partial \gamma}{\partial r'_k} d A' \nonumber \\ 
=& \iint_{\Omega(\bm{r}_c, \theta_c)} \frac{\partial \gamma}{\partial r'_i} d A' \nonumber \\
=& \iint_{\Omega(\bm{r}_c, \theta_c)} \left[ \nabla' \gamma \right]_i d A'.
\end{align}
Green's theorem is used in the calculation,
and $\varepsilon_{ij}$ is set as
\begin{align}
    \varepsilon_{ij} = \left\{ \begin{array}{ll} 1, & i = 1, j = 2, \\ -1, & i = 2, j = 1, \\ 0, & i = j, \end{array} \right.
\end{align}
where it should be noted that
\begin{align}
\varepsilon_{ij} \varepsilon_{jk} = - \delta_{ik},    
\end{align}
and
\begin{align}
\bm{a} \times \bm{b} = \varepsilon_{ij} a_i b_j
\end{align}
hold. Here, $\delta_{ij}$ is the Kronecker delta.
For the torque in Eq.~\eqref{torque_gen}, we obtain
\begin{align}
N =& \int_{\partial \Omega(\bm{r}_c, \theta_c)} (\bm{r}' - \bm{r}_{c}) \times \gamma(u(\bm{r}'))\bm{n}(\bm{r}') d\ell' \nonumber \\
=& \int_{\partial \Omega(\bm{r}_c, \theta_c)} \varepsilon_{ij} (r'_i - r_{c,i}) \gamma(u(\bm{r}')) n_j(\bm{r}') d\ell' \nonumber \\
=& \int_0^{2\pi} \gamma(u(\bm{q}(\theta'))) \varepsilon_{ij} (q_i(\theta') - r_{c,i})  \varepsilon_{jk} \frac{dq_k}{d\theta'} d\theta' \nonumber \\
=& -\int_0^{2\pi} \gamma(u(\bm{q}(\theta'))) (q_i(\theta') - r_{c,i})  \frac{dq_i}{d\theta'} d\theta' \nonumber \\
=& - \int_{\partial \Omega(\bm{r}_c, \theta_c)} \gamma(u(\bm{q})) (q_i - r_{c,i})  d q_i \nonumber \\ 
=& -  \iint_{\Omega(\bm{r}_c, \theta_c)} \varepsilon_{mi} \frac{\partial}{\partial r'_m}\left(\gamma(u(\bm{r}')) (r'_i - r_{c,i})  \right) d A' \nonumber \\ 
=& \iint_{\Omega(\bm{r}_c, \theta_c)}  \varepsilon_{mi} (r_i' - r_{c,i})\frac{\partial \gamma}{\partial r_m'} d A' \nonumber \\ 
=& \iint_{\Omega(\bm{r}_c, \theta_c)} \left[ (\bm{r}' - \bm{r}_c) \times \nabla' \gamma \right] d A'.
\end{align}

\section{Explicit Forms for $u_{i,j}^{(i)}$ and $u_{i,j}^{(o)}$ \label{appB}}

Here, we show the explicit forms for $u_{i,0}^{(i)}$ and $u_{i,0}^{(o)}$:
\begin{align}
    u_{1,0}^{(i)} =\frac{1}{\pi R^2}\frac{R}{2} \left[r \BK1{R} \BI0{r} - R \BK2{R} \BI1{r} \right] \cos\theta,
\end{align}
\begin{align}
    u_{1,0}^{(o)} =\frac{1}{\pi R^2}\frac{R}{2} \left[-r \BI1{R} \BK0{r} + R \BI2{R} \BK1{r} \right] \cos\theta,
\end{align}
\begin{align}
    u_{2,0}^{(i)} =& \frac{1}{\pi R^2} \frac{R^2}{32} \left[r^2 \left( \BK0{R} \BI0{r} - \BK2{R} \BI2{r} \right)  \right. \nonumber \\
    & \left. \quad - 2 \left( R \BK1{R} \BI0{r} - r \BK0{R} \BI1{r}\right) \right] \nonumber \\
    &+ \frac{1}{\pi R^2} \frac{R^2}{64} \left[ 2 r^2\left( \BK0{R} \BI0{r} - \BK2{R} \BI2{r} \right) \right. \nonumber \\
    &\left. \quad - rR \left( \BK1{R} \BI1{r} - \BK3{R} \BI3{r} \right) \right] \cos 2\theta,
\end{align}
\begin{align}
    u_{2,0}^{(o)} =& \frac{1}{\pi R^2} \frac{R^2}{32} \left[ r^2 \left( \BI0{R} \BK0{r} - \BI2{R} \BK2{r} \right) \right. \nonumber \\
    & \left. \quad + 2 \left(R \BI1{R} \BK0{r} - r\BI0{R} \BK1{r}\right) \right] \nonumber \\
    &+ \frac{1}{\pi R^2} \frac{R^2}{64} \left[ 2 r^2 \left( \BI0{R} \BK0{r} - \BI2{R} \BK2{r}\right) \right. \nonumber \\
    &\left. \quad  - rR \left(\BI1{R} \BK1{r}  - \BI3{R} \BK3{r}\right) \right] \cos 2\theta,
\end{align}
\begin{align}
    u_{3,0}^{(i)} =& \frac{1}{\pi R^2} \frac{R^2}{128} \left[ r(R^2 + r^2)(\BK2{R} \BI2{r} -\BK0{R} \BI0{r}) \right. \nonumber \\ & \left. \quad + 4 r \left( R \BK1{R} \BI0{r}  - r \BK0{R} \BI1{r} \right) \right] \cos \theta \nonumber \\
    &+ \frac{1}{\pi R^2} \frac{R^2}{1152} \left[ -3 r^3 \left( \BK0{R} \BI0{r} - \BK2{R} \BI2{r} \right)  \right. \nonumber \\ 
    &\left. \quad + 3 r^2R \left(\BK1{R} \BI1{r} - \BK3{R} \BI3{r} \right)   \right. \nonumber \\ & \left. \quad  -  rR^2 \left( \BK2{R} \BI2{r} - \BK4{R} \BI4{r} \right) \right] \cos 3\theta , \label{u30i}
\end{align}
\begin{align}
    u_{3,0}^{(o)} =& \frac{1}{\pi R^2} \frac{R^2}{128} \left[ r(R^2 + r^2)(\BI2{R} \BK2{r} -\BI0{R} \BK0{r}) \right. \nonumber \\ & \left. \quad - 4 r \left(R \BI1{R} \BK0{r} - r \BI0{R} \BK1{r}\right) \right] \cos \theta \nonumber \\
    &+ \frac{1}{\pi R^2} \frac{R^2}{1152} \left[ -3 r^3 \left( \BI0{R} \BK0{r} - \BI2{R} \BK2{r} \right) \right. \nonumber \\ 
    &\left. \quad + 3 r^2 R \left(\BI1{R} \BK1{r} - \BI3{R} \BK3{r} \right)   \right. \nonumber \\ & \left. \quad - rR^2 \left( \BI2{R} \BK2{r} - \BI4{R} \BK4{r} \right) \right] \cos 3\theta . \label{u30o}
\end{align}
Note that the above expressions are for the case with a circular camphor particle.

For the $n$-mode modification from a circle, we have the following explicit forms for $u_{i,1}^{(i)}$ and $u_{i,1}^{(o)}$:
\begin{align}
    u^{(i)}_{0,1} = \frac{1}{\pi R^2} \left[ R^2 \BK{n}{R} \BI{n}{r}\right] \cos n(\theta-\theta_c),
\end{align}
\begin{align}
    u^{(o)}_{0,1} = \frac{1}{\pi R^2} \left[R^2 \BI{n}{R} \BK{n}{r}\right] \cos n(\theta-\theta_c),
\end{align}
\begin{align}
    u^{(i)}_{1,1} =& \frac{1}{\pi R^2} \frac{R^2}{4} \left[ -r \BK{n}{R} \BI{n}{r} + R \BK{n-1}{R} \BI{n-1}{r} \right] \nonumber \\
    & \quad \times\cos((n-1)\theta - n\theta_c) \nonumber \\
    & + \frac{1}{\pi R^2} \frac{R^2}{4} \left[- r \BK{n}{R} \BI{n}{r} + R \BK{n+1}{R} \BI{n+1}{r}\right] \nonumber \\
    & \quad \times \cos ((n+1)\theta - n\theta_c),
\end{align}
\begin{align}
    u^{(o)}_{1,1} =& \frac{1}{\pi R^2} \frac{R^2}{4} \left[ -r \BI{n}{R} \BK{n}{r} + R \BI{n-1}{R} \BK{n-1}{r} \right] \nonumber \\
    & \quad \times\cos((n-1)\theta - n\theta_c) \nonumber \\
    & + \frac{1}{\pi R^2} \frac{R^2}{4} \left[- r \BI{n}{R} \BK{n}{r} + R \BI{n+1}{R} \BK{n+1}{r}\right] \nonumber \\
    & \quad \times \cos ((n+1)\theta - n\theta_c),
\end{align}
\begin{align}
    u^{(i)}_{2,1} =& \frac{1}{\pi R^2} \frac{R^2}{32} \left[ R^2 \BK{n-2}{R} \BI{n-2}{r}  - 2 rR \BK{n-1}{R} \BI{n-1}{r} \right. \nonumber \\ 
    &\left. \quad + r^2 \BK{n}{R} \BI{n}{r} \right] \cos((n-2)\theta - n\theta_c) \nonumber \\
    &+ \frac{1}{\pi R^2} \frac{R^2}{16} \left[ - \frac{n+1}{n} r R \BK{n-1}{R} \BI{n-1}{r} \right. \nonumber \\
    & \left. \quad + (r^2 + R^2) \BK{n}{R} \BI{n}{r} \right. \nonumber \\
   & \left. \quad - \frac{n-1}{n} rR \BK{n+1}{R} \BI{n+1}{r} \right]  \cos n(\theta - \theta_c) \nonumber \\
    &+ \frac{1}{\pi R^2} \frac{R^2}{32} \left[ r^2\BK{n}{R} \BI{n}{r} -2 rR \BK{n+1}{R} \BI{n+1}{r} \right. \nonumber \\
    & \left. \quad + R^2 \BK{n+2}{R} \BI{n+2}{r} \right] \cos ((n+2)\theta - n\theta_c),
\end{align}
\begin{align}
    u^{(o)}_{2,1} =& \frac{1}{\pi R^2} \frac{R^2}{32} \left[ R^2 \BI{n-2}{R} \BK{n-2}{r}  - 2 rR \BI{n-1}{R} \BK{n-1}{r} \right. \nonumber \\ 
    &\left. \quad + r^2 \BI{n}{R} \BK{n}{r} \right] \cos ((n-2)\theta - n\theta_c)  \nonumber \\
    &+ \frac{1}{\pi R^2} \frac{R^2}{16} \left[ -\frac{n+1}{n} r R \BI{n-1}{R} \BK{n-1}{r} \right. \nonumber \\
    & \left. \quad + (r^2 + R^2) \BI{n}{R} \BK{n}{r} \right. \nonumber \\
    & \left. \quad - \frac{n-1}{n} rR \BI{n+1}{R} \BK{n+1}{r} \right] \cos n(\theta - \theta_c) \nonumber \\
    &+ \frac{1}{\pi R^2} \frac{R^2}{32} \left[ r^2\BI{n}{R} \BK{n}{r}  -2 rR \BI{n+1}{R} \BK{n+1}{r} \right. \nonumber \\
    & \left. \quad  + R^2 \BI{n+2}{R} \BK{n+2}{r} \right] \cos ((n+2)\theta - n\theta_c),
\end{align}
\begin{align}
    u^{(i)}_{3,1} =& \frac{1}{\pi R^2} \frac{R^2}{384} \left[R^3 \BK{n-3}{R} \BI{n-3}{r}  \right. \nonumber \\
    & \left. \quad -3 r R^2 \BK{n-2}{R} \BI{n-2}{r} + 3 R r^2 \BK{n-1}{R} \BI{n-1}{r} \right. \nonumber \\
     & \left. \quad- r^3 \BK{n}{R}\BI{n}{r} \right]  \cos( (n-3)\theta + n\theta_c )  \nonumber \\
    & + \frac{1}{\pi R^2} \frac{R^2}{128} \left[ -\frac{n+1}{n-1} rR^2 \BK{n-2}{R} \BI{n-2}{r} \right. \nonumber \\
    & \left. \quad + R \left( \frac{2(n+1)}{n} r^2 + R^2 \right) \BK{n-1}{R} \BI{n-1}{r} \right. \nonumber \\
    & \left. \quad - r \left(r^2 + \frac{2(n-2)}{n-1} R^2   \right) \BK{n}{R} \BI{n}{r} \right. \nonumber \\
    & \left. \quad + r^2R \frac{n-2}{n} \BK{n+1}{R} \BI{n+1}{r} \right] \nonumber \\
    & \quad \times \cos((n-1)\theta - n\theta_c) \nonumber \\
    &+ \frac{1}{\pi R^2}\frac{R^2}{128} \left[ \frac{n+2}{n} r^2R \BK{n-1}{R} \BI{n-1}{r} \right. \nonumber \\ 
    & \left. \quad - r \left(r^2 + \frac{2(n+2)}{n+1} R^2\right) \BK{n}{R} \BI{n}{r} \right. \nonumber \\ 
    & \left. \quad + R \left( \frac{2(n-1)}{n} r^2 + R^2 \right) \BK{n+1}{R} \BI{n+1}{r} \right.\nonumber \\
    & \left. \quad - rR^2 \frac{n-1}{n+1} \BK{n+2}{R} \BI{n+2}{r} \right]  \nonumber \\ 
    & \quad \times \cos ((n+1)\theta- n\theta_c) \nonumber \\
    &+ \frac{1}{\pi R^2} \frac{R^2}{384} \left[ - r^3 \BK{n}{R} \BI{n}{r} \right. \nonumber \\
    & \left. \quad + 3 r^2R \BK{n+1}{R} \BI{n+1}{r} - 3 rR^2 \BK{n+2}{R} \BI{n+2}{r} \right. \nonumber \\
    & \left. \quad+ R^3 \BK{n+3}{R} \BI{n+3}{r} \right] \cos ((n+3) \theta - n\theta_c),
\end{align}
\begin{align}
    u^{(o)}_{3,1} =& \frac{1}{\pi R^2} \frac{R^2}{384} \left[R^3 \BI{n-3}{R} \BK{n-3}{r} \right. \nonumber \\
    & \left. \quad -3 r R^2 \BI{n-2}{R} \BK{n-2}{r} + 3 r^2R \BI{n-1}{R} \BK{n-1}{r} \right. \nonumber \\ 
    & \left. \quad - r^3 \BI{n}{R}\BK{n}{r} \right]\cos((n-3) \theta - n\theta_c )  \nonumber \\
    & + \frac{1}{\pi R^2} \frac{R^2}{128} \left[ -\frac{n+1}{n-1} rR^2 \BI{n-2}{R} \BK{n-2}{r} \right. \nonumber \\
    & \left. \quad + R \left( \frac{2(n+1)}{n} r^2 + R^2 \right) \BI{n-1}{R} \BK{n-1}{r} \right. \nonumber \\
    &\left. \quad - r \left(r^2 + \frac{2(n-2)}{n-1} R^2\right)\BI{n}{R} \BK{n}{r}  \right. \nonumber \\
    &\left. \quad + \frac{n-2}{n} r^2R \BI{n+1}{R} \BK{n+1}{r} \right] \nonumber \\
    & \quad \times \cos((n-1)\theta - n\theta_c) \nonumber \\
    &+ \frac{1}{\pi R^2}\frac{R^2}{128} \left[ \frac{n+2}{n} r^2R \BI{n-1}{R} \BK{n-1}{r} \right. \nonumber \\ 
    & \left. \quad - r \left( r^2 + \frac{2(n+2)}{n+1} R^2 \right) \BI{n}{R} \BK{n}{r} \right. \nonumber \\ 
    & \left.  \quad + R \left(\frac{2(n-1)}{n} r^2 + R^2 \right) \BI{n+1}{R} \BK{n+1}{r} \right. \nonumber \\
    & \left. \quad - \frac{n-1}{n+1} rR^2 \BI{n+2}{R} \BK{n+2}{r} \right]  \nonumber \\ 
    & \quad \times \cos ((n+1) \theta- n\theta_c) \nonumber \\
    &+ \frac{1}{\pi R^2} \frac{R^2}{384} \left[ - r^3 \BI{n}{R} \BK{n}{r} \right. \nonumber \\ 
    & \left. \quad + 3 r^2R \BI{n+1}{R} \BK{n+1}{r} - 3 rR^2 \BI{n+2}{R} \BK{n+2}{r} \right. \nonumber \\
    & \left. \quad + R^3 \BI{n+3}{R} \BK{n+3}{r} \right] \cos ((n+3)\theta - n\theta_c).
\end{align}

\end{document}